\documentclass{acm_sen_article}

\title{Role of Ontology Training to Software
Engineering Students
}
\author{Arvind W Kiwelekar\\
Department of Computer Engineering\\
Dr. B. A. Technological University\\
Lonere-402103 Raigad (MS) India\\
awk@dbatu.ac.in
}

\begin{document}

\maketitle

\begin{abstract}
Students of software engineering struggle to develop a
systems perspective because most of the software engineering
methodologies focus on developing a particular aspect of a system.
Lack of unified coverage to the topic of
systems modeling is identified as the root cause behind this  problem.
The paper explains the role of ontology in
building systems perspective.
A case for the necessity of ontology training as a means to ovrcome
this problem is prsented. 
  The course content for a typical
course on ontology  is also described in the paper.

\end{abstract}

\section{Introduction}
The skill of systems modeling is one of the necessary skills
for executing  software engineering projects beside other
generic skills such as project management and communication skills.
% Systems modeling is acthe ability to represent physical
% and socio-technical systems 
One of the shortcomings of the curricula followed
for computer engineering programmes spcially in Indian Universities 
is that the topic of
systems modeling is either taught in a fragmented way or
under-emphasized. Some of the effects of the lack of unified
coverage to the topic of systems modeling are that students
fail to develop a systemic view of an engineering application
and they also fail to relate theoretical concepts to real world
objects.

\section{LIMITATIONS OF SE CURRICULA}
Table 1 depicts the coverage of the topic of systems modeling included in a typical computer engineering curriculum.
These courses mainly focus on different approaches to systems modeling and to explain various process models for
building software systems. Few problems associated with
such kinds of course curricula are:
(i)\textbf{ Fragmented treatment to the topic of systems
modeling.} Systems modeling related topics such as data
modeling, function modeling, agent modeling, kinds of sysems are covered in different courses. The effect of such
isolated treatment to these topics reflect in students failure
to grasp the commonalities among similar concepts and their
failure to relate the concepts studied in different courses. For
example, students struggle to relate the similar concepts of
Finite Automata and State Diagram that are taught in the
subjects of TOC and OOAD.
(ii) \textbf{Emphasis on symbolic expression over semantic
expression} While teaching the subjects like TOC, OOAD,
DM there is more emphasis on writing efficient symbolic
expression and on syntax of a particular formalism such as
state machines, first order logic and programming languages.
Often neglecting that symbols stand for the objects from
reality. This emphasis get reflected in student's failure to
relate theoretical concepts to real world knowledge. Most
of the students always label states with symbols such as $s_{1}$,
$s_{2}$ ... and transitions with $e_{1}$, $e_{2}$ .... when 
students are asked to  draw a state transition diagram for a library book. 
More meaningful terms such as $onTheRack$, $issued$,
$newArrival$ are rarely thought over.
(iii) \textbf{Lack of explicit coverage to the notions of System
and }Time The notions of system, kinds of systems, emergent properties of systems, environment, and time which are
central to the task of systems modeling are unaddressed in
present undergraduate computer engineering curricula. Due
to this students face difficulties in understanding more ad-
vanced courses such as Temporal Logic, and systems archi-
tecting.
(iv) \textbf{Abstract mathematical concepts are taught before concrete concepts} The abstract mathematical con-
cepts of graph theory, relations, algebraic structures, queu-
ing theory are taught prior to more concrete SE concepts
and they are explained through non-software engineering
applications. Students find easier to learn abstract concepts
when they are taught in terms of familiar and more concrete
applications. As a result building mathematical models of
software systems appear as a major challenge.
\section{ONTOLOGY AND SYSTEM MODELING}
The term Ontology is primarily used in two different ways.
Ontology as a philosophical discipline\cite{Bunge} studies various
kinds of objects found in reality. From the information sci-
ence point of view, the term ontology formally specifies con-
cepts found in a particular application domain \cite{Chand}. In this
paper, the term ontology is used in both senses to explain
the role of ontologies in addressing the issues raised in the
last section.
\begin{table*}[t]
\begin{center}
\begin{tabular}{|p{1.5in}|p{1.5in}|p{2.5in}|}
\hline 
Systems Modeling Tasks & Course  & Typical Coverage \\ \hline
 
Data Modeling & Data base Management (DBMS) & Entity and Relationship modeling, Query Languages,
                                         Transaction management, Architecture of DBMS \\ \hline
Object Modeling & Object Oriented Analysis and De sign (OOAD) & Object-Oriented Programming, UML notation, Data
                Abstraction, Inheritance, Ploymorphism, and Design
                           Patterns \\ \hline 
Function Modeling & Information System Analysis  and Design (ISAD) &  Information gathering, Structured analysis, Func-
  tional decomposition. \\ \hline 
Agent Modeling  & Artificial Intelligence (AI) & Problem solving techniques, Knowledge representa-
                                            tion, Logical reasoning. \\ \hline 
Mathematical Modeling &  Discrete Mathematics (DM) and  Theory of Computations (TOC) & First Order logic and Finite State Automata/State Modeling \\ \hline 
Software Systems Development &  Software Engineering (SE)  & Engineering processes of software systems such as re-
 quirements analysis, design, coding, maintenance etc. \\ \hline 
\end{tabular}
\caption{Systems Modeling Topics coverage in a Computer Engineering Curriculum} \label{t1}
\end{center}
\end{table*}

As a philosophical discipline, ontologies formally define the
notions of things, properties, events, processes, agents, in-
tentions and relationships among them thus providing a sin-
gle unified framework to the concepts employed in systems
modeling. In field of information systems modeling, ontolo-
gies have been applied as a foundational framework to eval-
uate the effectiveness of modeling languages.
The topic of ontology is currently partially covered in AI
courses with an objective to represent domain knowledge in
AI applications. In such applications, a domain ontology
is one of the core components. The current trend of build-
ing intelligent applications and devices further justifies the
necessity of ontology training to software engineers.

Ontologies provide a set of categories that can be employed
to describe a particular application domain, analyze an ap-
plication domain, and express the facts about domain ele-
ments. During the task of ontological analysis, the focus of a
modeler is directed on semantic issues of model elements and
relating model elements to the kinds of objects found in the
reality. During many systems analysis task (eg. CRC card,
DFD), modelers are trained to think in terms of language
constructs. During ontological analysis, a modeler engages
in a systemic thinking rather than on capturing some frag-
mented views of a system.
The philosophical discipline of Ontology makes least number
of assumptions and it defines the concepts that are taken for
granted in many specialized engineering courses on modeling
languages, mathematics and logics. Hence the notion of sys-
tem, kinds of systems, relationships of systemic properties
with its component, time, and time dependent properties
are explicitly defined in an ontology. Thus bridging the loose
connections existing between modeling abstractions and do-
main knowledge.
The abstract concepts of finite automata, logical formu-
lae, language abstractions provide system modelers a set of
mechanisms to represent the objects from reality. Ontologies
provide the content to employ these mechanisms during the
task of systems modeling. The existing curricula adopt the
mechanisms first approach while teaching systems modeling
topics. The student’s comprehension of abstract concepts
can be improved if these concepts are introduced through
concrete ontological categories.

\section{ONTOLGY:COURSE CONTENT}
The course on ontology \cite{ot} can be taught during third or fourth
semester of the undergraduate engineering programme. A
typical course on ontology may include the topics such as :
(i) Ontology as a philosophical discipline (ii) Upper level on-
tologies (iii) Ontological categories from Aristotle and Bunge
ontology. (iv) Role of Ontology in Information Systems
Modeling. (v) Ontology Specification Languages: Descrip-
tions Logic, OWL. ( vi) Building Domain Ontologies. (vii)
Applications of ontologies in engineering applications.
The effect of ontology training in understanding scientific
concepts has been assessed in other disciplines such as Physics
and Medicine. Similar experiments can be conducted before
introducing a full fledged course on ontology in computer
engineering programme.
\bibliographystyle{abbrv}
\bibliography{paper}

\end{document}